\documentclass{ws-ijmpb}
\usepackage{graphicx,amsbsy,amsmath,bbm}
\newcommand{\sma}{A}
\newcommand{\smb}{B}
\newcommand{\smc}{C}
\newcommand{\smab}{{AB}}
\newcommand{\smac}{{AC}}
\newcommand{\smbc}{{BC}}
\newcommand{\KPsi}{|\boldsymbol{T}\rangle}
\newcommand{\bmsigma}{\boldsymbol{\sigma}}
\newcommand{\bmSigma}{\boldsymbol{\Sigma}}
\newcommand{\sA}{A}
\newcommand{\sB}{B}
\newcommand{\sC}{C}
\newcommand{\sT}{T}
\newcommand{\bmA}{{\boldsymbol A}}
\newcommand{\bmB}{{\boldsymbol B}}
\newcommand{\bmC}{{\boldsymbol C}}
\newcommand{\bmD}{{\boldsymbol D}}
\newcommand{\bmO}{{\boldsymbol O}}

\newcommand{\bmT}{{\boldsymbol T}}
\newcommand{\ii}{\mathbbm{1}}
\newcommand{\pp}{\mathbb{P}}
\newcommand{\bmOmega}{\boldsymbol \Omega}
\newcommand{\bmomega}{\boldsymbol \omega}
\newcommand{\Sent}{{\mathcal S}}
\newcommand{\Disc}{{\mathcal D}}
\newcommand{\hO}{a}
\newcommand{\mL}{{\cal L}}
\begin{document}
\markboth{S.~Olivares and M.~G.~A.~Paris}
{The balance of quantum correlations for a class....}
\catchline{}{}{}{}{}
\title{THE BALANCE OF QUANTUM CORRELATIONS FOR A CLASS OF FEASIBLE 
TRIPARTITE CONTINUOUS VARIABLE STATES}
\author{STEFANO OLIVARES}
\address{Dipartimento di Fisica, Universit\`a degli Studi di Milano,
I-20133 Milano, Italy\\
CNISM UdR Milano Statale, I-20133 Milano, Italy\\
stefano.olivares@fisica.unimi.it}
\author{MATTEO G.~A.~PARIS}
\address{Dipartimento di Fisica, Universit\`a degli Studi di Milano,
I-20133 Milano, Italy\\
CNISM UdR Milano Statale, I-20133 Milano, Italy\\
matteo.paris@fisica.unimi.it}
\maketitle
\begin{history}
\received{Day Month Year}
\revised{Day Month Year}
\end{history}
\begin{abstract}
  We address the balance of quantum correlations for continuous
  variable (CV) states. In particular, we consider a class of feasible
  tripartite CV pure states and explicitly prove two Koashi-Winter-like
  conservation laws involving Gaussian entanglement of formation,
  Gaussian quantum discord and sub-system Von Neumann entropies. We also address
  the class of tripartite CV mixed states resulting from the
  propagation in a noisy environment, and discuss how the previous
  equalities evolve into inequalities.
\end{abstract}
\section{Introduction}\label{s:intro}
Multipartite quantum correlations in continuous variable (CV) systems are
valuable resources for quantum technology
\cite{sch86,nap04,ade04,bra05,wol06,wee11}.  Fully inseparable
three-mode Gaussian states \cite{pvl00,Gie01,pvl03} have been proposed to
realize cloning at distance  \cite{pvl01,OliPar}, and experimental schemes to
generate multimode CV entangled states have been already suggested and
demonstrated \cite{fur98,Zha02,Jin03,Aok03,Glo03,OL04,3M04}.
\par
More recently, more general quantum correlations in bipartite Gaussian
states have been analyzed: Gaussian quantum discord has been
introduced \cite{GQD10,Ade10} and experimentally investigated
\cite{expD1,expD2,expD3}.
For discrete variables several interesting relations among 
the different measures of quantum correlations have been introduced
\cite{Koa04,Kay09,Paw09,See10,Pra11,Glg11,Fan11,Ren11} with the aim of 
clarifying their different meaning and role \cite{Mod12}.

The dimension of the Hilbert space does not appear to play a crucial
role and, indeed, balance of correlations for CV systems has been
derived in terms of the Renyi entropy \cite{Ade12}. In addition, 
balance of correlations in quantum measurements, and during the 
propagation in noisy channels, has been discussed both for discrete 
and continuous variables \cite{Bru11,Gio11,Cam11,Ges12,Chu12,Str12}.
\par
Motivated by these results, in this paper we investigate whether 
some form of conservation laws for quantum correlations may be 
established for continuous variable systems and Gaussian measures of 
correlations. In particular, we consider a specific class of feasible 
pure CV tripartite states, and analyze in details the balance of Gaussian 
correlations
among the bipartite subsystems. We found that Koashi-Winter-like
relations, involving entanglement of formation, quantum discord and
system entropies, may be generalized to continuous variables using Gaussian
measures of quantum correlations, at least for the
class of feasible states that we took into account. The results are
encouraging enough to suggest a direct experimental verification and to
foster investigations of correlation balance in more general Gaussian
states.
Our results may also taken as an argument in favor of
the conjecture that Gaussian quantum discord is {\em the} quantum discord for 
Gaussian states \cite{All12} and contribute to the ongoing
discussion on the nature of quantum correlations in continuous variable
systems \cite{pvi12}. 
\par
The paper is structure as follows. In the next Section we introduce the
class of feasible tripartite Gaussian states we are going to consider. 
In Section \ref{s:bal} we present the conservation laws for quantum
correlations, whereas in Section \ref{s:noi} we consider the propagation
in a noisy environment, and discuss how the previous 
equalities evolve into inequalities. Section \ref{s:out} closes the
paper with some concluding remarks.
\section{A class of feasible tripartite Gaussian states} 
\label{s:tri}
Let us consider three modes of a bosonic field, coupled through the 
interaction Hamiltonian 
\begin{equation}
H =\gamma_1 a^\dag c^{\dag} 
+ \gamma_2 b^{\dag} c + h.c.
\label{H3}\,.
\end{equation}
The Hamiltonian $H$ describes, e.g,  two interlinked bilinear
interactions taking place among three modes of the radiation field. In
fact, it has been studied long ago \cite{And70,SLu74} for the description
of parametric processes in $\chi^{(2)}$ media, assuming a suitable
configuration satisfying phase-matching  conditions.  Interaction
schemes described by $H$ has been been realized using a single
$\chi^{(2)}$ nonlinear crystal where the two interactions take place
simultaneously \cite{OL04}.  The three-mode entanglement
generated by $H$ may be indeed exploited to realize optimal cloning at 
distance of coherent states \cite{3M04}. Analogue interaction schemes
may be realized in condensate systems \cite{tlb03,tlb04}.
\par
The Hamiltonian in Eq. (\ref{H3}) admits the constant of motion
$$K(t) \equiv N_\sma(t) - N_\smb(t) - N_\smc(t) \equiv K (0),$$ 
where $N_j(t)$ denotes the average number of photons in the $j$-th mode. 
Starting from the vacuum $|{\bf 0\rangle}\equiv |0\rangle_1 \otimes |0\rangle_2
\otimes |0\rangle_3$ we have $K(0)=0$ and thus $N_\sma(t)=N_\smb(t)+N_\smc(t)$ 
$\forall t$. The expressions for
$N_j(t)$ can be obtained by the Heisenberg evolution of the field
operators; upon introducing $\Omega = \sqrt{|\gamma_2|^2 -|\gamma_1|^2}$ we have
\begin{eqnarray}
N_\sma = N_\smb+N_\smc \;, \nonumber \qquad
N_\smb = \frac{|\gamma_1|^2 |\gamma_2|^2}{\Omega ^4}
\left[\cos{\Omega t}-1 \right]^2 \;, \qquad
N_\smc = \frac{|\gamma_1|^2}{\Omega ^2} \sin^2(\Omega t) \nonumber
\end{eqnarray}
The evolved state $\KPsi  =U_t |{\bf 0}\rangle$ reads as follows
\begin{equation}
\KPsi  = \frac{1}{\sqrt{1+N_\sma}} \sum_{nm}
\left(\frac{N_\smb}{1+N_\sma}\right)^{n/2}
\left(\frac{N_\smc}{1+N_\sma}\right)^{m/2}
\sqrt{\frac{(n+m)!}{n! m!}}\: |n+m,n,m\rangle
\label{state}\;,
\end{equation}
where $U_t=\exp\left(-iH t\right)$ is the evolution operator. 
The states $\KPsi$ are Gaussian
states, since they are generated from the vacuum by a bilinear
Hamiltonian.  Upon introducing the canonical operators $q_k$ and $p_k$,
$k=A,B,C$, and the vector
$R=(q_\sma,p_\sma,q_\smb,p_\smb,q_\smc,p_\smc)^{\sT}$, it is
straightforward to prove that the mean values of the canonical operators
are zero $\langle R \rangle \equiv \langle \bmT|R | \bmT \rangle=0$, 
whereas the covariance matrix (CM), whose elements are given
by $$\bmSigma_{hk}=\frac{1}{2}\langle\{R_k,R_h\}\rangle-\langle
R_k\rangle\langle R_h\rangle,$$  can be written in the 
following block form:
\begin{align}
\label{CovPsi}
\bmSigma_{\bmT} &= \left(
\begin{array}{ccc}
\bmD_\sma & \bmO_{\sma \smb} & \bmO_{\sma \smc} \\
\bmO_{\sma \smb} & \bmD_\smb & \bmO_{\smb \smc} \\
\bmO_{\sma \smc} & \bmO_{\smb \smc} & \bmD_\smc
\end{array}
\right)\,.
\end{align}
where:
\begin{subequations}
\begin{align}
\label{CovPsiAux}
\bmD_\sma&=\left(N_\sA+\frac12\right)\,\ii
&\bmD_\smb&=\left(N_\sB+\frac12\right)\,\ii
&\bmD_\smc&=\left(N_\sC+\frac12\right)\,\ii \\[2ex]
\bmO_{\sma \smb} &= \sqrt{N_\sB(N_\sA+1)}\,\pp
&\bmO_{\sma \smc} &= \sqrt{N_\sC(N_\sA+1)}\,\pp
&\bmO_{\smc \smb} &= \sqrt{N_\sB N_\sC}\,\ii \;,
\end{align}
\end{subequations}
with $N_{\sma} = N_{\smb} + N_{\smc}$, 
$\ii={\rm Diag}(1,1)$ and $\pp={\rm   Diag}(1,-1)$.  
Since partial trace is a Gaussian operation the single-mode partial traces:
$$
\varrho_\sma=\hbox{Tr}_{\smb\smc}\Big[|\bmT
\rangle\langle\bmT|\Big],\quad
\varrho_\smb=\hbox{Tr}_{\sma\smc}\Big[|\bmT
\rangle\langle\bmT|\Big],\quad
\varrho_\smc=\hbox{Tr}_{\sma\smb}\Big[|\bmT \rangle\langle\bmT|\Big],
$$
and the two-mode ones
$$
\varrho_{\sma\smb}=\hbox{Tr}_{\smc}\Big[|\bmT
\rangle\langle\bmT|\Big],\quad
\varrho_{\sma\smc}=\hbox{Tr}_{\smb}\Big[|\bmT
\rangle\langle\bmT|\Big],\quad
\varrho_{\smb\smc}=\hbox{Tr}_{\sma}\Big[|\bmT \rangle\langle\bmT|\Big],
$$
are Gaussian states as well. The corresponding CM may be
obtained by dropping the corresponding lines in $\bmSigma_{\bmT}$. The
single-mode partial traces thus have diagonal CM $\bmsigma_k =
\bmD_k$, corresponding to thermal states:
\begin{equation}
\nu (N_k) = \sum_p \frac{N_k^p}{(N_k+1)^{p+1}}\, |p\rangle\langle p|\,,\:\: 
k=A,B,C,
\end{equation}
and the corresponding von Neumann entropies are given by 
$$\Sent_k =f\left(\sqrt{\det \bmsigma_k}\right)
= f\left(N_k+\frac12\right),$$
where: 
\begin{equation}
f(x)= \left(x + \frac12\right) \ln\left(x +\frac12\right) -
\left(x -\frac12\right) \ln\left(x - \frac12\right) \quad
\hbox{for} \:\: x>\frac12\,.
\end{equation}
The CM of the two-mode partial traces are given by
\begin{align}
\bmsigma_{\smab}= \left(
\begin{array}{cc}
\bmD_\sma & \bmO_{\sma \smb}  \\
\bmO_{\sma \smb} & \bmD_\smb  \\
\end{array}
\right)\,,\quad
\bmsigma_{\smac}= \left(
\begin{array}{cc}
\bmD_\sma & \bmO_{\sma \smc}  \\
\bmO_{\sma \smc} & \bmD_\smb  \\
\end{array}
\right)\,,\quad
\bmsigma_{\smbc}= \left(
\begin{array}{cc}
\bmD_\smb & \bmO_{\smb \smc}  \\
\bmO_{\smb \smc} & \bmD_\smc  \\
\end{array}
\right)\,.
\end{align}
and are already in the so called standard block form (to which 
any CM may be brought by means of local symplectic transformations) 
$${\bmsigma}= \left(\begin{array}{cc} \bmA & \bmC \\ \bmC & \bmB
\end{array}\right ) \quad 
\bmA=\hbox{Diag}(a,a),\:\bmB=\hbox{Diag}(b,b), \:  
\bmC=\hbox{Diag}(c_{1},c_{2})\,.$$ 
For $\bmsigma_{\sma\smb}$  and $\bmsigma_{\sma\smc}$ we have  
$c_1=-c_2$ whereas for $\bmsigma_{\smb\smc}$ we have  
$c_1=c_2$. This means that $\varrho_\smab$ and $\varrho_\smac$ are
squeezed thermal states (STS) of the form:
\begin{align}
\varrho_\smab &= S(r)\, \nu (N_\smc) \otimes |0\rangle\langle 0|\, S^\dag (r),
\quad \sinh^2 r = \frac{N_\smb}{1+N_\smc}\,, \\
\varrho_\smac &= S(r)\, \nu (N_\smb) \otimes |0\rangle\langle 0|\, S^\dag (r),
\quad \sinh^2 r = \frac{N_\smc}{1+N_\smb}\,,
\end{align}
respectively, $S(r)=\exp\{r (a^\dag b^\dag - ab)\}$ being the two-mode squeezing
operator, whereas $\varrho_\smbc$ corresponds to a mixed 
thermal state (MTS) that may equivalently expressed as (recall that $N_A = N_B+N_C$):
\begin{align}
\varrho_\smbc &= U(\phi)\, \nu (N_A) \otimes |0\rangle\langle
0|\, U^\dag (\phi)\,, \quad \cos^2 \phi = \frac{N_\smb}{N_\smb+N_\smc}\,, \\
\varrho_\smbc &= U(\phi)\, |0\rangle\langle 0| \otimes \nu
(N_A) \, U^\dag (\phi)\,,
\quad \cos^2 \phi = \frac{N_\smc}{N_\smb+N_\smc}\,,
\end{align}
where $U(\phi)=\exp\{\phi (a^\dag b - ab^\dag)\}$ is a bilinear mixing
operator describing e.g. the action of a beam splitter.
\section{Balance of correlations}\label{s:bal}
In order to investigate the correlations between the modes of the state
$\KPsi$, we introduce the symplectic eigenvalues $\lambda_\pm^{(hk)}$ of
the two mode state $\varrho_{hk}$ and the minimum symplectic eigenvalue
$\tilde{\lambda}_{-}^{(hk)}$ of its partial transpose, which may be
obtained as:\cite{OliTut} \begin{align}
\lambda_\pm^{(hk)} &= 2^{-1/2}\,\sqrt{I_h+I_k+2I_{hk} \pm 
\sqrt{(I_h+I_k+2I_{hk})^2-4 J_{hk}}}\,,
\notag \\ 
\tilde{\lambda}_{-}^{(hk)} &= 2^{-1/2}\,\sqrt{I_h+I_k-2I_{hk} 
- \sqrt{(I_h+I_k-2I_{hk})^2-4 J_{hk}}}\,,
\end{align}
respectively, where 
$$I_{k}=\det \bmD_k,\quad I_{hk}=\det \bmO_{hk},\quad J_{hk}=\det \bmsigma_{hk}
\,.$$ 
Note that:
\begin{align}\label{vN:entropy}
\Sent_k = f\left( \sqrt{I_k}\right)\,.
\end{align}
whereas the von Neumann entropies of the two-mode partial traces are
given by
\begin{align}\label{vN:entropy2}
\Sent_{hk} = f(\lambda_+^{(hk)})+ f(\lambda_-^{(hk)})\,.
\end{align}
The positivity of the state is equivalent to
$\lambda_\pm^{(hk)} \geq \frac12$ and the positivity of the 
partially transposed state to $\tilde{\lambda}_{-}^{(hk)} \geq \frac12$. 
The bipartite Gaussian state $\varrho_{hk}$ is thus entangled iff 
$0 \leq \tilde{\lambda}_{-}^{(hk)} < \frac12$.  In our case we have
\begin{align}\label{our:case}
\lambda_-^{(hk)}= \frac12
\quad \mbox{and} \quad 
\lambda_+^{(hk)}= \frac12 + N_{j}
\end{align}
and thus the von Neumann entropy of any two-mode partial trace 
equals the von Neumann entropy of the remaining (single) mode, 
in formula:
\begin{align} \label{onetwo}
\Sent_{hk} = \Sent_j
\end{align}
where, in both Eqs. (\ref{our:case}) and (\ref{onetwo}), we have 
$h\ne k\ne j$ and $h,k,j=A,B,C$.
\par
For the symplectic eigenvalues of the partial transposes we have
\begin{align}
\tilde\lambda_-^{\sma\smb}&=
\frac{1}{\sqrt{2}}\Bigg[
\left(N_\sma+\frac{1}{
      2}\right)^2+2 N_\smb
         (N_\sma+1)+\left(N_\smb+\frac{1}{2}\right)^2
\notag \\
&\mbox{}\hspace{3cm}-(N_\sma+N_\smb+1) \sqrt{4 N_\smb (N_\sma+1)+N_\smc^2}
\Bigg]^{1/2} \:,  \\[2ex]
\tilde\lambda_-^{\sma\smc} &=
\tilde\lambda_-^{\sma\smb} (N_\sma \leftrightarrow N_\smb) \:, \\[2ex]
\tilde\lambda_-^{\smb\smc} &=
\Bigg[ (N_\smb - N_\smc)^2 + N_\smb+ N_\smc+ \frac12
- \left|N_\smb-N_\smc\right|  \sqrt{\left(N_\sma+1\right){}^2-4 N_\smb
N_\smc} \Bigg]^{1/2} \:,
\end{align}
from which it is easy to see that for any value of $N_\smb$ and $N_\smc$
we have 
$$0 \leq \tilde\lambda_-^{(\sma\smb)} \leq \frac12 \qquad
0\leq \tilde\lambda_-^{(\sma\smc)} \leq \frac12 
\qquad \tilde\lambda_-^{(\smb\smc)} \geq \frac12\,,
$$
i.e., $\varrho_{\sma\smb}$ and $\varrho_{\sma\smc}$ are entangled
states whereas $\varrho_{\smb\smc}$ is separable.
If $0\le\tilde{\lambda}_{-}^{(hk)} < \frac12$, i.e., $\varrho_{hk}$ 
is entangled, then the Gaussian entanglement of formation (EoF) is given 
by $$E_{hk}(\varrho)=h(y_{hk})\,,$$ where:\cite{eof03,Marian}
\begin{equation}\label{y:EoF}
y_{hk}=\frac{(\sqrt{I_h}+\sqrt{I_k})(\sqrt{I_hI_k} - |I_{hk}| + 1/4) -
2 \sqrt{|I_{hk}| \tilde{J}_{hk}}}{(\sqrt{I_h}+\sqrt{I_k})^2-4 |I_{hk}|}\,,
\end{equation}
with $\tilde{J}_{hk} = \det (\bmsigma_{hk} + \frac12 \bmOmega)$, and 
the symplectic form $\bmOmega$ given by
$$\bmOmega = \bmomega\oplus\bmomega \qquad \bmomega = 
\left(\begin{array}{cc} 0 & 1 \\ -1 & 0 \end{array}\right)\,.$$
In our case, $\varrho_{AB}$ and $\varrho_{AC}$ are entangled and we have:
\begin{equation}\label{y:EoF:hk}
y_\smab = y_{BA} = \frac12 + \frac{N_\smb}{1+N_\smc}\,,
\quad y_\smac= y_{CA} = \frac12 + \frac{N_\smc}{1+N_\smb}\,,
\end{equation}
respectively.  On the other hand, being the state $\varrho_{BC}$ separable, 
$y_{BC}= y_{CB} =\frac12$ and thus its EoF is zero.
\par
For MTS and STS the Gaussian quantum discords 
$\Disc^\rightarrow(\varrho_{hk})\equiv \Disc_{hk}$ and
$\Disc^\leftarrow(\varrho_{hk})\equiv \Disc_{kh}$ (measurements
performed on the first and second mode respectively) may be expressed
as:\cite{GQD10} 
\begin{subequations}\label{discord}
\begin{align}
\Disc_{hk} &=  f\left(\sqrt{I_{h}}\right)+
f\left(\sqrt{I_{k}}-\frac{2 |I_{hk}|}{1+2 \sqrt{I_{h}}}\right)
-f\left(\lambda_{-}^{(hk)}\right)-f\left(\lambda_{+}^{(hk)}\right)\,, \\
\Disc_{kh} &=  f\left(\sqrt{I_k}\right)+
f\left(\sqrt{I_{h}}-\frac{2 |I_{hk}|}{1+2 \sqrt{I_{k}}}\right)
-f\left(\lambda_{-}^{(hk)}\right)-f\left(\lambda_{+}^{(hk)}\right)\,.
\end{align}
\end{subequations}
\par
Starting from Eqs.~(\ref{discord}) we can write a conservation law for
the quantum correlations of the two-modes states $\varrho_{hk}$. At
first, we rewrite Eqs.~(\ref{discord}) as follows:
\begin{subequations}\label{discord:balance}
\begin{align}
\Disc_{hk}  + h\left(\lambda_{+}^{(hk)}\right) &= \Sent_h +
f\left(\sqrt{I_{k}}-\frac{2 |I_{hk}|}{1+2 \sqrt{I_{h}}}\right)
- f\left(\lambda_{-}^{(hk)}\right)\,, \\
\Disc_{kh} + h\left(\lambda_{+}^{(hk)}\right) &=  \Sent_k+
f\left(\sqrt{I_{h}}-\frac{2 |I_{hk}|}{1+2 \sqrt{I_{k}}}\right)
- f\left(\lambda_{-}^{(hk)}\right)\,,
\end{align}
\end{subequations}
where we used Eq.~(\ref{vN:entropy}). Besides, for the state (\ref{state}), 
Eqs.~(\ref{our:case}) say that $f\left(\lambda_{-}^{(hk)}\right) = 0$ and
$f\left(\lambda_{+}^{(hk)}\right) = \Sent_j$, $h\ne k\ne j$, where we
used the identity $\lambda_{+}^{(hk)} = \sqrt{I_j}$. Furthermore,
it is straightforward to verify that: 
\begin{equation}
\sqrt{I_{h}}-\frac{2 |I_{hk}|}{1+2 \sqrt{I_{k}}} =
y_{hj} \quad (h\ne k\ne j)
\end{equation}
with $y_{hj}$ given in Eqs.~(\ref{y:EoF:hk}).
Summarizing, for the state $| \bmT \rangle$ in Eq.~(\ref{state}), 
Eqs.~(\ref{discord:balance}) leads to the following conservation laws 
\begin{subequations}\label{discord:balance:ideal}
\begin{align}
\Disc_{hk} + \Sent_j &= \Sent_h + E_{kj}\,, \label{discord:balance:a} \\
\Disc_{hk} + \Sent_{hk} &= \Sent_h + E_{kj}\,, \label{discord:balance:kw}
\end{align}
\end{subequations}
with $h\ne k\ne j$ and $h,k,j=A,B,C$.
\par
The above equalities are explicitly showing that
Koashi-Winter-like conservation laws for quantum correlations may 
be written for continuous variable systems using Gaussian measures for quantum
correlations, at least for the class of feasible states
described by Eq. (\ref{state}). Eq. (\ref{discord:balance:kw}) is the
direct counterpart of the balance of correlations obtained for discrete 
variables, and thus we may expect, or conjecture, that it is of general
validity for CV systems. On the other hand, 
Eq. (\ref{discord:balance:a}) is due to the specific properties of 
the state $\KPsi$.  Notice also that the validity of  Eq. 
(\ref{discord:balance:kw}) may represent an argument in favor of
the conjecture that Gaussian quantum discord is {\em the} quantum discord for 
Gaussian states \cite{All12}. Finally, we notice that both Eqs. 
(\ref{discord:balance:a}) and (\ref{discord:balance:kw})
represent an independent check of
the formula for the EoF in the case of non symmetric states
\cite{Marian}. 
\section{Balance of correlations in the presence of noise}\label{s:noi}
In this section we investigate how, and to which extent, the
conservation laws (\ref{discord:balance:ideal}) addressed
in the previous section are modified by the propagation of the state in
noisy channels. We assume that the three modes evolve through three 
identical uncorrelated noisy channels and that the Markovian
approximation is valid. The propagation is thus described
by the following Master equation
\begin{equation}\label{ME}
\dot{\varrho}_t = \frac{\gamma}{2}\sum_{k=1,2,3}\left\{
(N_{\rm th}+1) \mL [a_k] + N_{\rm th} \mL [a_k^\dag]\right\}\,\varrho_t\,,
\end{equation}
where $\varrho_t$ is the density matrix of the tripartite system
described by the field operators $a_1=a$, $a_2=b$ and $a_3=c$, $\mL
[\hO]\varrho_t=2 \hO\varrho_t \hO^{\dag}-\hO^{\dag}\hO\varrho_t -
\varrho_t \hO^{\dag} \hO$ is the Lindblad superoperator, $\gamma$ is
the overall damping rate, while $N_{\rm th}$ represents the effective
number of photons of the noisy channels \cite{nap04}. Upon preparing
the three modes in the initial state $\KPsi$, the Gaussian nature is
preserved during the evolution and the evolved CM $\bmSigma_t$ can be
written as:\cite{OliTut}
\begin{equation}\label{evol:CM}
  \bmSigma_\tau = e^{-\tau}\,\bmSigma_{\bmT} +
(1-e^{-\tau})\, \bmSigma_{\infty}
\end{equation}
where we introduced $\tau=\gamma t$, $\bmSigma_{\bmT}$ is the CM of the
initial state given in Eq.~(\ref{CovPsi}) and $\bmSigma_{\infty} =
\frac12(1+2N_{\rm th}) \ii_{4}$ is the asymptotic CM, $\ii_{4}$ being 
the $4\times 4$ identity matrix.  
\par
\begin{figure}[h!]
\includegraphics[width=0.99\textwidth]{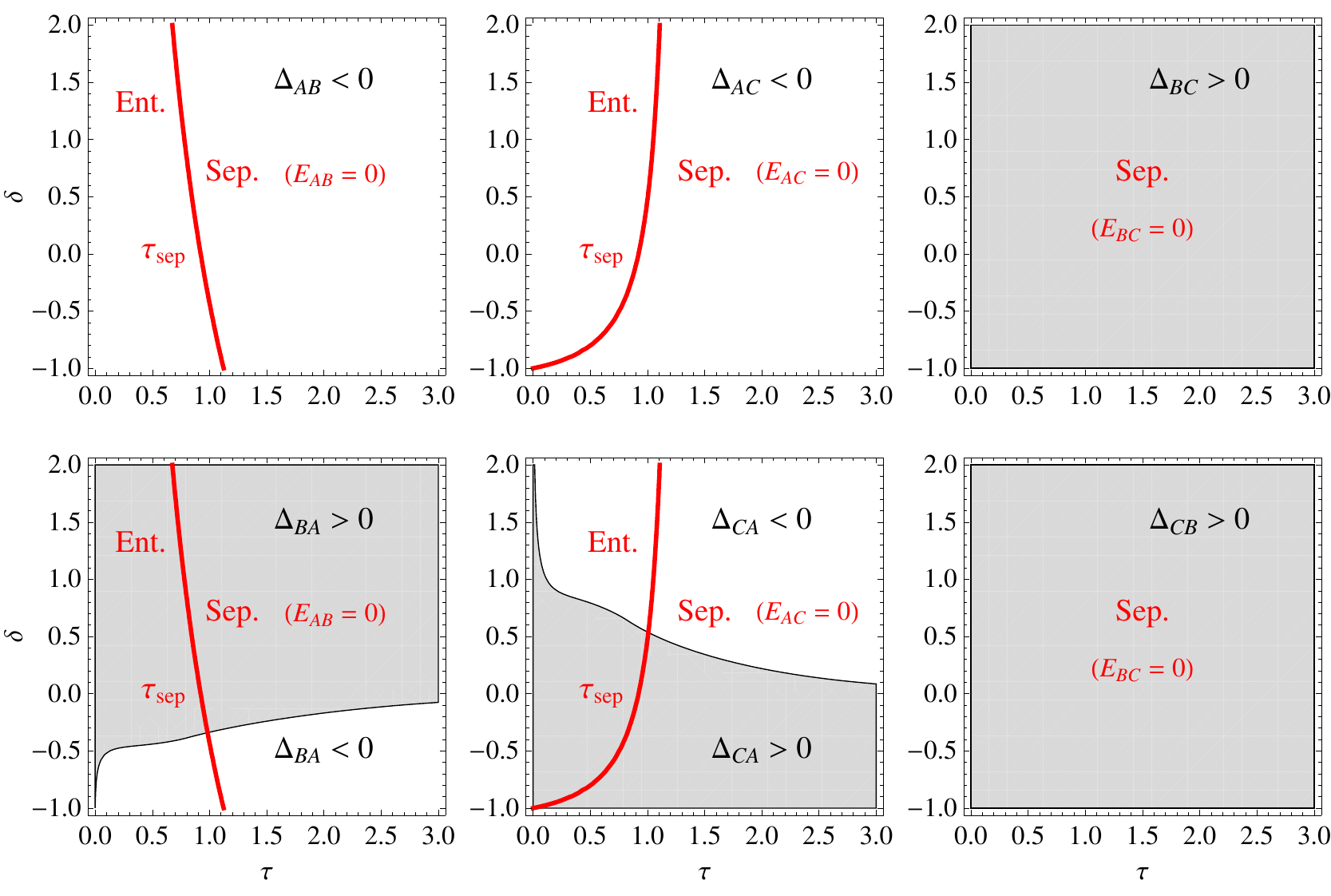}
\caption{\label{f:balance:noise}(Color online) Balance of the
  correlations in the presence of dissipation and thermal
  noise. Region plots of $\Delta_{hk}$, $h\ne k$ with $h,k=A,B,C$, as
  functions of $\delta = N_{B}-N_{C}$ of $\tau = \gamma t$. We set
  $N_{B}=1$ and $N_{\rm th}=0.2$. Gray and white regions refer to the
  corresponding $\Delta_{hk} > 0$ and $\Delta_{hk} < 0$,
  respectively. The red line represents the separability time
  $\tau_{\rm sep}$: if $\tau \ge \tau_{\rm sep}$ then
  $\tilde{\lambda}^{(hk)}_{-} \ge \frac12$ and the state
  $\varrho_{hk}(t)$ is no longer entangled (Ent.) and becomes
  separable (Sep.), i.e., $E_{hk} = 0$. See the text for details.}
\end{figure}
\par
By using the evolved CM (\ref{evol:CM}) and its time-dependent local
symplectic invariants, we can calculate the quantities appearing in
Eqs.~(\ref{discord:balance}). In particular, now one has:
\begin{equation}
f\left(\lambda_{-}^{(hk)}\right) \ge 0\,
\quad \hbox{and} \quad
f\left(\lambda_{+}^{(hk)}\right) \le \Sent_{j}
\end{equation}
and, in turn, the equalities (\ref{discord:balance:ideal}) changes
into inequalities.  The analytic expressions
of the evolved conservation laws may be evaluated analytically, but they 
are quite clumsy and are not reported here explicitly.
\par
In order to study the evolution of the law (\ref{discord:balance:a})
we define the following function:
\begin{equation}\label{Delta} \Delta_{hk}(\tau) =
  \left[\Disc_{hk}(\tau) + \Sent_j(\tau)\right] - \left[\Sent_h(\tau)
    + E_{kj}(\tau)\right]
\end{equation}
where all the involved quantities are calculated starting from the
results of the previous section but with the evolved state
$\varrho_{t}$.
As an example, Fig.~\ref{f:balance:noise} is a region-plot
$\Delta_{hk}(\tau)$ as functions of $\tau$ and $\delta=N_B-N_C$ for
given $N_B$ and $N_{\rm th}$: depending on the values of the involved
parameters, $\Delta_{hk}(\tau)$ can be positive or negative. In
particular, $\Delta_{AB}$ and $\Delta_{AC}$ are always
negative, $\Delta_{BA}$ whereas d $\Delta_{CA}$ can change the sign
(remarkably, this holds true also for $N_{\rm th}=0$). For the case of
symmetric states. i.e., $\delta = 0$, one can write the following set
of inequalities:
\begin{subequations}
\begin{align}
\Disc_{AB} + \Sent_C &\le \Sent_A\,,
&\Disc_{BA} + \Sent_C &\ge \Sent_B + E_{AC}\,, \\
\Disc_{AC} + \Sent_B &\le \Sent_A\,,
&\Disc_{CA} + \Sent_B &\ge \Sent_C + E_{AB}\,, \\
\Disc_{BC} + \Sent_A &\ge \Sent_B + E_{CA}\,,
&\Disc_{CB} + \Sent_A &\ge \Sent_C + E_{BA}\,,
\end{align}
\end{subequations}
where we used $E_{BC} = E_{CB} = 0$.
\begin{figure}[h!]
\includegraphics[width=0.99\textwidth]{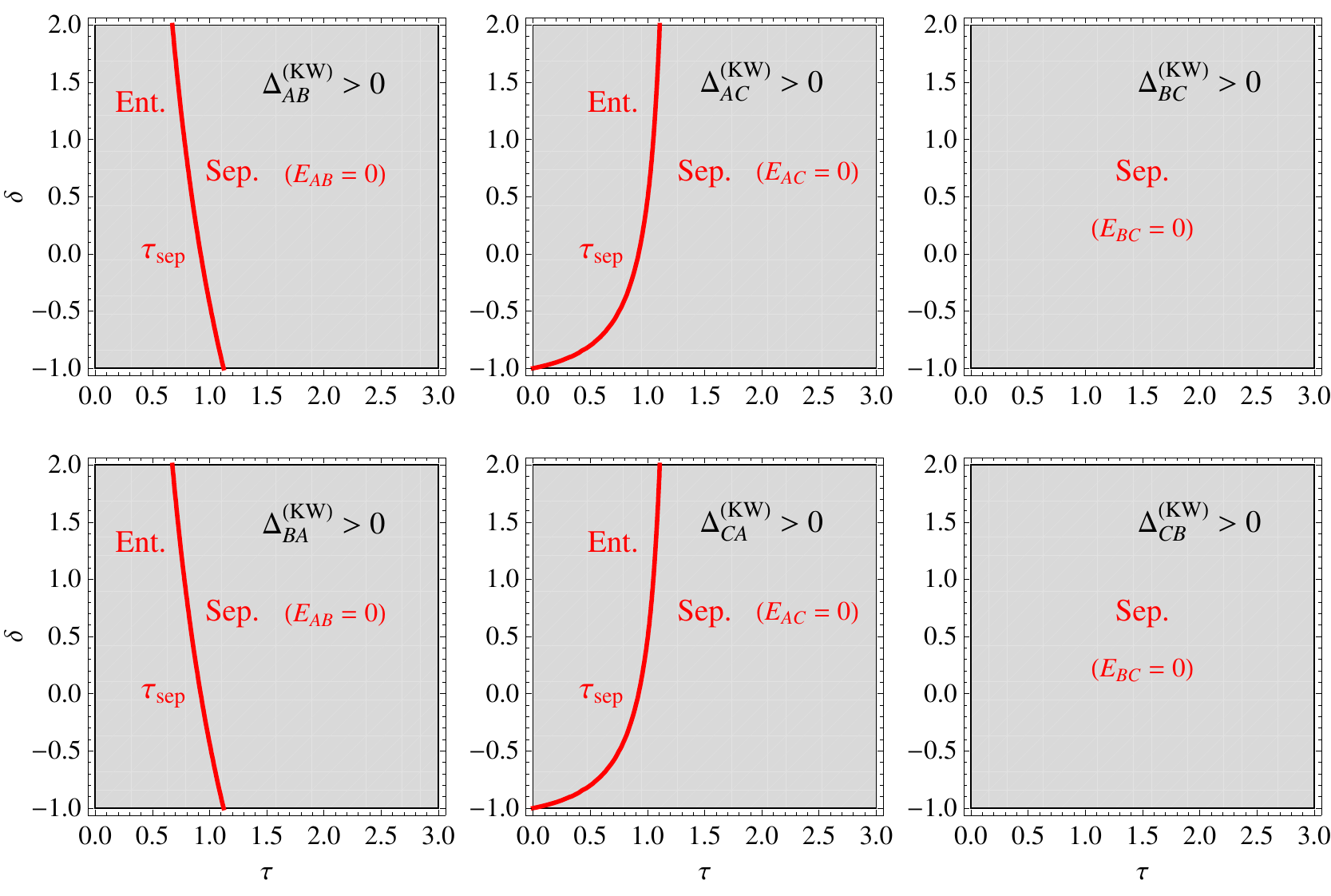}
\caption{\label{f:balance:noise:kw}(Color online) Balance of the
  correlations in the presence of dissipation and thermal
  noise. Region plots of $\Delta_{hk}^{\rm (KW)}$, $h\ne k$ with
  $h,k=A,B,C$, as functions of $\delta = N_{B}-N_{C}$ of $\tau =
  \gamma t$, for the same choice of the other involved parameters as
  in Fig.~\ref{f:balance:noise}. Note that one always has $\Delta_{hk}
  > 0$ (gray regions). The red line still represents the separability
  time $\tau_{\rm sep}$. See the text for details.}
\end{figure}
\par
Analogously, to address the evolution of the conservation law
(\ref{discord:balance:kw}), we introduce the function:
\begin{equation}\label{Delta:KW} \Delta_{hk}^{\rm (KW)}(\tau) =
  \left[\Disc_{hk}(\tau) + \Sent_{hk}(\tau)\right] - \left[\Sent_h(\tau)
    + E_{kj}(\tau)\right]\,.
\end{equation}
As one can see in Fig.~\ref{f:balance:noise:kw}, where we plot
$\Delta_{hk}^{\rm(KW)}(\tau)$ for the same choice of parameters as in
Fig.~\ref{f:balance:noise}, now one finds the following inequality
holding for all the bipartitions, and any value of the interaction time,
\begin{equation} \label{last}
\Disc_{hk}(\tau) + \Sent_{hk}(\tau) \ge \Sent_h(\tau) + E_{kj}(\tau)\,,
\end{equation}
$h\ne k\ne j$ and $h,k,j=A,B,C$.
Ineq. (\ref{last}) generalizes to CV and Gaussian measures of
correlation, the
inequality discussed in \cite{Fan11} for the discrete case.
\section{Conclusions}\label{s:out}
In conclusion, we have proved that the balance of correlations originally 
investigated for three-qubit systems, involving entanglement of formation, 
quantum discord and single-system entropies is valid also for a feasible 
class of tripartite Gaussian CV states, upon using Gaussian measures of
quantum correlations.
Furthermore, in the presence of dissipation and thermal noise, the
balance turns into inequalities between the previous quantities,
depending on the actual values of the involved parameters. The results
are encouraging enough to suggest a direct experimental verification and
to foster investigations of correlation balance in more general Gaussian
states.
\section*{Acknowledgments}
This work has been supported by MIUR (FIRB ``LiCHIS'' - RBFR10YQ3H).
MGAP thanks Kavan Modi, Gerardo Adesso, Natalia Korolkova, 
Laura Mazzola, Sabrina Maniscalso, 
Paolo Giorda and Ruggero Vasile for discussions.


\end{document}